# INFORMATION RETRIEVAL SYSTEM FOR SILT'E LANGUAGE

## Using BM25 weighting

**Abdulmalik Johar [1]**

*Lecturer, Department of information system, computing and informatics college, Wolkite University, Ethiopia*

---***---

**Abstract -** *The main aim of an information retrieval system is to extract appropriate information from an enormous collection of data based on user's need. The basic concept of the information retrieval system is that when a user sends out a query, the system would try to generate a list of related documents ranked in order, according to their degree of relevance. Digital unstructured Silt'e text documents increase from time to time. The growth of digital text information makes the utilization and access of the right information difficult. Thus, developing an information retrieval system for Silt'e language allows searching and retrieving relevant documents that satisfy information need of users. In this research, we design probabilistic information retrieval system for Silt'e language. The system has both indexing and searching part was created. In these modules, different text operations such as tokenization, stemming, stop word removal and synonym is included.*

## 1. INTRODUCTION

Information retrieval is a set of activities that represent, store and ultimately obtain information based on a specific need [1]. IR deals with the representation, storage, organization of, and access to information items [2]. Nowadays, information is everywhere that helps people to make the correct decision at the right time. As the written information becomes large in size and digital documents easily available electronically, it would be difficult to retrieve relevant documents among the accumulated document collections.

### 1.1. Statement of the Problem

There are more than 80 languages in Ethiopia. Silt'e is one of the zonal languages in the southern Nations, Nationalities and people's region of Ethiopia. According to the Central Statistics Authority (2007:80), the number of Silt'e speakers is 751,159. The census report reveals that 47,097 people or 6.28% of the total population are urban or semi-urban inhabitants [3] [4]. It is the South-Ethio-Semitic language family and uses Saba Ge'ez (Ethiopic script). A lot of peoples speak it in many parts of Ethiopia. Elementary schools in Silt'e zone give all subjects in Silt'e from grade one up to four. After grade four, it is given as one subject up to preparatory level and takes it as grade ten national exams. The Silt'e zone administration decided to use the langue as a working language since 2014 [10].

Digital unstructured Silt'e text documents in government organizations and schools increase from time to time in the Silt'e zone. The growth of digital text information makes the utilization and access of the right information difficult. Therefore, an information retrieval system needed to store, organize and helps to access Silt'e digital text information

### 1.2. Literature Review

**The Writing System of Silt'e Language**

The accurate period that the Ge'ez character set was used for Silt'e writing system is not known. However, since the 1980s, Silt'e has been written in the Ge'ez alphabet, or Ethiopic script, writing system, originally developed for the now extinct Ge'ez language and most known today in its use for Amharic and Tigrigna [9]. Portion of the Ge'ez (Ethiopic) alphabets and their phonetic order are shown in the following table.

| 1st order | 2st order | 3st order | 4st order | 5st order | 6st order | 7st order |
|---|---|---|---|---|---|---|
| ሀ ha | ሁ hu | ሂ hi | ሃ haa | ሄ he | ህ hi | ሆ ho |
| ለ le | ሉ lu | ሊ li | ላ la | ሌ le | ል li | ሎ lo |
| መ me | ሙ mu | ሚ mi | ማ ma | ሜ me | ም mi | ሞ mo |
| ረ re | ሩ ru | ሪ ri | ራ ra | ሬ re | ር ri | ሮ ro |
| ሰ se | ሱ su | ሲ si | ሳ sa | ሴ se | ስ si | ሶ so |

*Table 1. Phonetic order of Ge'ez alphabets*

A Probabilistic Information Retrieval System for Tigrinya evolved via Atalay [8] to test the prototype system evolved, 300 Tigrinya documents and 10 queries had been used to check the technique. The prototype is developed utilizing Python 3.2.2 programming language. The system registered, after stemming and pseudo-significance feedback, an average precision 69.1%, recalls 90%, and F-measure 74.4%. This result is achieved without controlling the issue of synonyms and polysemy of terms that exist in Tigrinya textual content

Jemal Z. developed Silt'e Text Information Retrieval System Based on Vector Space Model. He utilized Apache Lucene Solr to design the prototype. To test the prototype system developed 100 Silt'e archive corpus. Those experimentations performed using the prototype that was developed without invoking stemming algorithm scored 81.4% mean average precision. This end result is accomplished without controlling the issue of synonyms and polysemy of phrases that exist in Silt'e textual document [5].

Developing information retrieval system has a great impact on the accessing of information in a particular language. We have seen in the literature there are different information retrieval systems designing approaches and each model has its own strength and limitation. To enhance the correctness and efficiency of the retrieved information from the corpus we have to select appropriate information retrieval system development approaches

## 2. METHODOLOGY

### 2.1. Silt'e IR System Design and Architecture

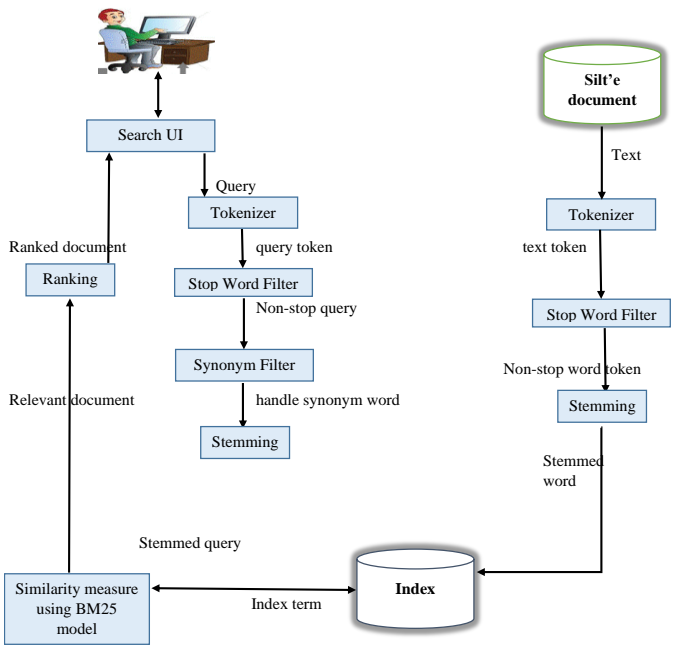

*Figure 1. Silt'e information retrieval system architecture*

### 2.2. Tokenization

Tokenizer accepts the text stream, processes the characters, and produces a sequence of tokens. It can break the text stream based on characters such as whitespace or symbols [1]. Consider the statement "ሉላሉሌ ዪንጄ ሉባምቸነ ‹‹ፋገ››፣ ‹‹፣አረሺ››፣ ‹‹ቡደ››፣››" the tokens after implementing standard tokenizer splitting the sentence are "ሉላሉሌ", " ዪንጄ", "ሉባምቸነ","ፋገ","አረሺ" and "ቡደ". Therefore the purpose of tokenization is to break down a document collection into tokens so that it is easy to match with tokens in the query. As a result, it increases the relevancy of a document to a query.

### 2.3. Stop Word Removal

Not all terms found in the document are equally important to represent documents they exist in. Some terms are most common in documents. Therefore, removing those terms, which are not used to identify some portions of the document collection, is important. Such terms are removed based on two methods. The first method is to remove high frequent terms by counting the number of occurrences (frequency). The second method is using stop word list for the language [6]. In this research, we used the second method. The stop word list is adopted from Muzeyn K. [7]. below shows sample stop word list used in this research

| አብተእብታይ | ወይ | ቢትላይም | ገጌ |
| አብተቴ | ታሌ | አሎነ | ሀነግነ |
| አብታይ | ናር | ሂነሚ | እነይ |
| አብተቴ | ናርት | በሁነትነሙ | ሱርዋ |
| እሊ | ናሩ | ሁኖተኒመዋ | ገናሚ |
| እሊ | ብቾ | ኣታይ | ሂነኩምነገ |
| እነይ | ለገነ | ሀነይ | ኡሀ |
| እነይ | ገነ | ሁኖ | ያሽ |
| እቲ | ሁልምክ | ገና | ዮዬ |
| እቲ | ለሳድባድክ | ዮልስ | በዮ |
| እቢ | ለሰባድሽ | ሉላሉሌ | ዋ |
| እቢ | ለሰባድኑም | በሉሌ | ለገነገናም |

*Table 2. sample stop word list*

### 2.4. Performance Evaluation

Once information retrieval system is designed and developed it is essential to carry out its evaluation. Evaluate an IR system is to measure how well the system meets the information needs of the users. Without appropriate retrieval evaluation, cannot determine how well the IR system is performing. There are different techniques of evaluation of the performance of a system based its goal. In fact, the primary goal of an IR system is to retrieve all the documents which are relevant to a user query while retrieving as few non-relevant documents as possible [2].

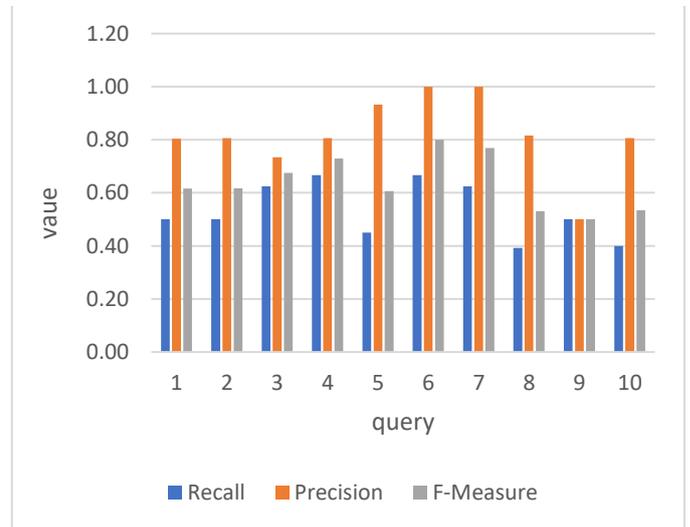

*Chart -1. The average recall, f-measure and precision of the system without stemming and synonym*

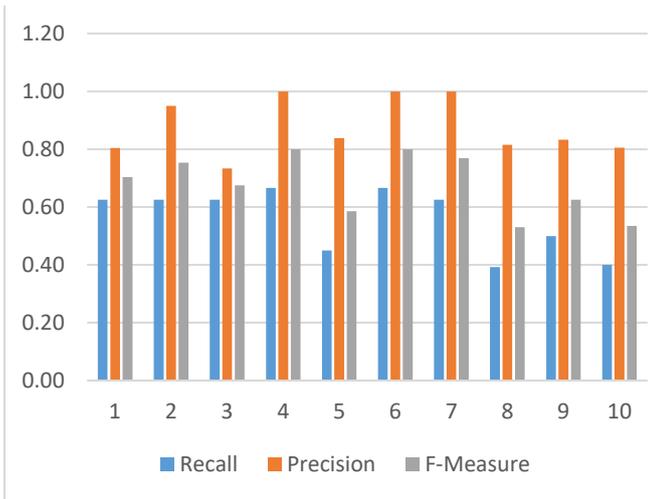

*Chart -2. The average recall, f-measure and precision of the system with stemming and synonym*

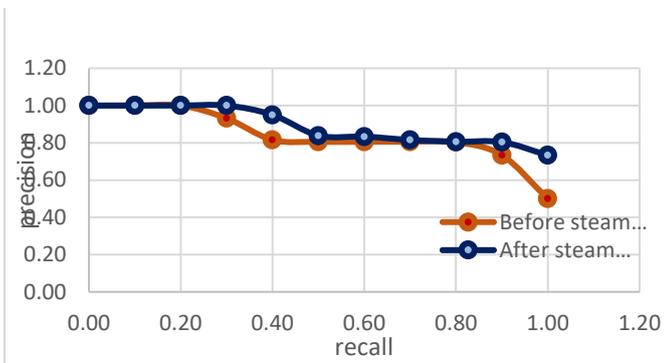

*Chart -3. Average Recall-Precision curve before and after steam and synonym*

## 3. CONCLUSIONS

Text retrieval system is very important for retrieval of textual documents. The study tries to develop probabilistic IR system for Silt'e language. The developed prototype has two parts: indexing and searching. The indexing part of the work involves tokenization, stop word removal, synonym and steaming using Apache solr filtering factor. For seaming use steaming dictionary supported by solr. The system developed has also searching components. The main parts are query pre-processing, similarity measurement and ranking the document. The retrieval model used for the study is one of the probabilistic models BM25. The experimentation performed using the prototype that was developed scored 88% mean average precision. To achieve better result Apache Solr allows us to add our own module to process Silt'e text.

Generally, the study shows good results in developing text information retrieval system for Silt'e language. However, further exploration and study should be done to improve and produce a better performance by increasing the corpus size in the language.